# Near optimal efficient decoding from pooled data

MAX HAHN-KLIMROTH, TU Dortmund University, Germany

NOELA MÜLLER, Eindhoven University of Technology, the Netherlands

Consider $n$ items, each of which is characterised by one of $d+1$ possible features in $\{0, \ldots, d\}$. We study the inference task of learning these types by queries on subsets, or pools, of the items that only reveal a form of coarsened information on the features - in our case, the sum of all the features in the pool. This is a realistic scenario in situations where one has memory or technical constraints in the data collection process, or where the data is subject to anonymisation. Related prominent problems are the quantitative group testing problem, of which it is a generalisation, as well as the compressed sensing problem, of which it is a special case. In the present article, we are interested in the minimum number of queries needed to efficiently infer the labels, if one of the features, say 0, is dominant in the sense that the number $k$ of non-zero features among the items is much smaller than $n$. It is known that in this case, all features can be recovered in exponential time by using no more than $O(k)$ queries. However, so far, all *efficient* inference algorithms required at least $\Omega(k \ln n)$ queries, and it was unknown whether this gap is artificial or of a fundamental nature. Here we show that indeed, the previous gap between the information-theoretic and computational bounds is not inherent to the problem by providing an efficient algorithm that succeeds with high probability and employs no more than $O(k)$ measurements. This also solves a long standing open question for the quantitative group testing problem.

## 1 INTRODUCTION

Imagine a population of $n$ items, each of which is characterised by one of finitely many distinct features, such as a class label in an object detection task, an age group, gender, blood type. We are interested in inferring these types by using only a small number of coarse measurements on subgroups of the items. This problem, as introduced by [26], is known as the *pooled data problem*. While the general framework is of relevance in many practical situations, a particularly prominent and topical example of the pooled data problem is the quantitative group testing problem [6, 12, 13, 16]. In this case, the population is split into two types, which we interpret as *healthy* and *defective*, and the goal is to identify which individuals are the defective ones, by using tests that can only provide the total number of defectives in the pooled subgroup. Other applications of the pooled data problem include DNA screening [23], traffic monitoring [25], machine learning [19, 20] and signal recovery [21].

In the following, we will assume that the labels of the items are chosen uniformly from a suitable set. Since the information on the features increases with the number of queries we ask on them, a natural and important question concerns the minimum number of queries that are needed to successfully identify all the labels with high probability[1] over the choice of the labels. To address it, one can derive both upper and lower bounds: A sequence $m_0$ such that for $m \geq m_0$, the probability of making an error does not tend to one is called an *information-theoretic lower bound*. On the other hand, a sequence $m_0$ such that for $m \geq m_0$, *exponential time* algorithms like exhaustive search are guaranteed to recover all features w.h.p. is called an *information-theoretic upper bound*.

Having obtained information-theoretic bounds and thus identified a regime where inference is possible, a second important question is concerned with the minimal $m$ for which all features can be recovered *efficiently* (in polynomial time).

---

[1]With high probability (w.h.p.) means with probability tending to 1 as $n \to \infty$.

Our article addresses this second question in the case where one of the features is dominant. For this setting, we provide an efficient algorithm that w.h.p. infers all labels correctly, while using no more than $O(k)$ queries. This algorithm is the first one to match the information-theoretically optimal order of queries. In the special case of quantitative group testing, this result resolves the long standing open question whether efficient information-theoretically optimal inference is possible. We continue to describe the precise model and related results.

### 1.1 Model and terminology

Consider $n$ items $x_1, \ldots, x_n$, each of which is assigned a label $\sigma_i := \sigma(x_i) \in \{0, 1, 2, \ldots, d\}$. The vector $\boldsymbol{\sigma} = (\sigma_1, \ldots, \sigma_n)$ of item-labels constitutes the *ground-truth* or *signal* that we aim to infer by performing $m$ queries on subsets $a_1, \ldots, a_m \in 2^{\{1,\ldots,n\}}$ of the items, which we call *pools*. There is freedom both in the choice of the pools $a_1, \ldots, a_m$ as well as in the nature of the queries. In the design of the pools, we restrict ourselves to the *non-adaptive* setting, where all $m$ pools have to be constructed before conducting any queries. This assumption is particularly relevant in situations where the evaluation of the queries dominates the running time of the inference algorithm - a specific example is feature detection in machine learning, in which feeding an image into a deep neural network on a GPU is time-consuming. With respect to the queries, we consider the *additive model*, where for each $i = 1, \ldots, m$, the total *weight* $\hat{\sigma}_i := \sum_{j \in a_i} \sigma_j$ of pool $i$ is measured. In particular, this model is a natural extension of the quantitative group testing problem, where the measurement $\hat{\sigma}_i$ corresponds to the number of defectives in the $i$-th pool. On the other hand, it reveals less information on the labels than other commonly studied variants of the problem, where one measures a histogram of the frequencies of each label within the given pool (see [8, 22]).

It is well-known that the presence of a dominant label, say 0, is a distinguishing feature in the theoretical analysis of the model. Denote by $k_0, \ldots, k_d$ the numbers of items with label $0, \ldots, d$, respectively, and by $k := \sum_{i=1}^{d} k_i$ the total number of non-zero labels. We call the fraction $k/n$ of non-zero entries of $\boldsymbol{\sigma}$ the *sparsity* of the model. If $k/n \to \alpha \in (0, 1)$ for $n \to \infty$, the problem is called *linear pooled data problem*, while the *sublinear pooled data problem* considers $k \sim n^\theta$ for some $\theta \in (0, 1)$. In this article, we study the sublinear regime. One reason behind the interest in this regime is its practicability. In particular, according to Heaps' Law of Epidemiology [1], early spreads of epidemics follow this model. Furthermore, pooled measurements are applied in feature detection tasks in machine learning [19, 20, 27]. In this setting, rare occurrences of specific labels need to be found on images in which most images only contain background information (corresponding to label 0).

Within the sublinear setting, we assume that $k_0, \ldots, k_d$ are known a priori, for example through empirical findings, and that for each label $i = 1, \ldots, d$, there exists $\varepsilon_i = \Theta(1)$ such that $k_i = \varepsilon_i k$. This model choice is analogous to parametrisations in the linear regime of the recent study by El Alaoui et al. [8]. Finally, we perform an average-case analysis of the problem and from here on, we will thus assume that the ground-truth $\boldsymbol{\sigma}$ is chosen uniformly at random among all vectors having exactly $k_i$ entries of type $i$, where $i = 0, \ldots, d$.

In the next two subsections, we briefly review some previous results and then give an overview of our main results. After fixing some notation in Section 2, Sections 3 and 4 describe our pooling scheme and inference algorithm.

### 1.2 Information-theoretic and computational bounds

*Information-theoretic lower bounds.* A meaningful information-theoretic lower bound $m_{\text{count}}$ can be obtained through a simple counting argument: For fixed $k_0, \ldots, k_d$, there are $\binom{n}{k_0, k_1, \ldots, k_d}$ different values that the ground-truth $\boldsymbol{\sigma}$ can take. On the other hand, each query outputs a value between 0 and $W := \sum_{j=1}^{d} jk_j$. For unambiguous inference to be possible with high probability, we thus need $(W + 1)^{m_{\text{count}}} \gtrsim \binom{n}{k_0, k_1, \ldots, k_d}$, where $\gtrsim$ means inequality with respect to the leading exponential order. In our setting, where $k = n^\theta$ for $\theta \in (0, 1)$, this argument has the consequence that for fewer than $m_{\text{count}} = \Omega(k)$ pools, there is no hope to successfully reconstruct $\sigma$ w.h.p. In the special case of the quantitative group testing problem, Djackov [6] provides the explicit information-theoretic lower bound $m_{\text{QGT}} \geq 2 \frac{1-\theta}{\theta} k$.

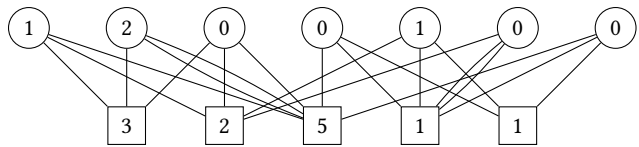

Fig. 1. Graphical representation of a pooling scheme: Here, the $n = 7$ items are represented by circles, while the $m = 5$ pools are represented by squares. Edges between items and pools are present whenever an item is an element of the corresponding pool.

*Information-theoretic upper bounds.* Using exhaustive search, Grebinski & Kucherov show that

$$m_{\text{GK}} = 4W \ln\left(\frac{n}{W} + 1\right) \ln^{-1}(W) = O(k)$$

pools suffice to reconstruct $\sigma$ w.h.p. [13]. Moreover, in the special case of the quantitative group testing problem, the leading constant was further reduced by a factor of 2 in independent works of [12] and [10]. Since these results asymptotically match the information-theoretic lower bound, the pooled data problem can be regarded as almost understood from an information-theoretic point of view. However, with respect to efficient algorithms, the picture is a different one.

*Efficient Algorithms.* Natural candidates as efficient inference algorithms for the pooled data problem are algorithms that also work for the more general sparse compressed sensing problem. The literature on this topic is vast, and we refrain from reviewing it here. Many of these articles are based on improved versions of the Basis Pursuit algorithm, with the prominent contributions of [3, 7, 11]. All those approaches have in common that they require $\Omega(k \ln n)$ measurements in the sublinear regime.

In the extensively studied quantitative group testing problem, more specialised algorithms exist, see e.g. [5, 10, 12, 15], and one might hope that taking into account the specific structure of the problem yields some improvement in the order of the pools needed. However, this is not the case, and also these ideas have failed to beat the lower bound of $\Omega(k \ln n)$. Thus, when comparing what is information-theoretically achievable and what is efficiently achievable, in all cases, we are faced with a multiplicative gap of order $\ln n$.

### 1.3 Our contribution

In this article, we propose and analyse an efficient algorithm that reconstructs $\sigma$ correctly with no more than $O(k)$ measurements w.h.p., thereby overcoming the multiplicative $\ln n$ gap between the current information-theoretic and algorithmic bounds. Our algorithm makes use of a random, almost regular pool design that is based on the so-called *spatial coupling*-technique from coding theory [17, 18]. A similar technique has recently been applied to the binary group testing problem [5]. The algorithm itself is then based on a thresholding idea. In the quantitative group testing setting, its performance matches the information-theoretic lower bound $m_{\text{QGT}}$ up to a moderate constant. Overall, we achieve the following main result.

THEOREM 1.1 (POOLED DATA PROBLEM). *Let $d \in \mathbb{N}$ be fixed, $k = n^\theta$ for $\theta \in (0, 1)$, $\varepsilon_1(n), \ldots, \varepsilon_d(n) \in (0, 1)$ s.t. $\sum_{i=1}^d \varepsilon_i(n) = 1$. Moreover, let $\sigma \in \{0, 1, \ldots, d\}^n$ be uniformly chosen among all vectors having exactly $\varepsilon_i k$ entries of value $i$ for $i = 1, \ldots, d$. Then for each $\delta > 0$, there are a polynomial-time construction of a pooling scheme and a polynomial-time algorithm that recovers $\sigma$ w.h.p. under the additive model using no more than*

$$m_{PD} = (8 + \delta) \frac{1 + \sqrt{\theta}}{1 - \sqrt{\theta}} \left( \sum_{w=1}^d w^2 \varepsilon_w \right) \frac{1 - \theta}{\theta} k$$

*pools.*

In particular, for the *quantitative group testing problem*, Theorem 1.1 implies the following.

COROLLARY 1.2 (QUANTITATIVE GROUP TESTING). *Let $k = n^\theta$ for $\theta \in (0, 1)$ and $\sigma \in \{0, 1\}^n$ be a uniformly chosen vector of Hamming weight $k$. Then for any $\delta > 0$, there is a polynomial-time construction of a testing scheme coming with a polynomial-time algorithm that recovers $\sigma$ w.h.p. using no more than $m_{SC} = (8 + \delta) \frac{1+\sqrt{\theta}}{1-\sqrt{\theta}} \frac{1-\theta}{\theta} k$ tests.*

### 1.4 Related problems

Extensions of the quantitative group testing problem ($d = 1$) have been studied since the 1960's by, among others, Erdős and Rényi [9], Djackov [6] and Shapiro [24]. El Alaoui et al. [8] and [26] introduce a variant where $\sigma$ is is a vector in $\{0, 1, \ldots, d\}^n$ and queries output the numbers of items of each label within the pool, while Bshouty [2] studies the *coin weighting problem*, where every query returns the sum of contained labels. Clearly, any algorithm that succeeds within the setting of Bshouty [2] can also be used within the framework of El Alaoui et al. [8], Wang et al. [26]. Finally, the pooled data problem can be seen as a special case of the *compressed sensing problem* [3, 7, 11], which generally asks for recovery of a high-dimensional signal $\sigma \in \mathbb{R}^n$ from a small number of linear measurements on its components.

## 2 MODEL

### 2.1 Getting started

Recall that we aim to infer the labels $\sigma_1, \ldots, \sigma_n \in \{0, 1, \ldots, d\}$ of $n$ items $x_1, \ldots, x_n$ by measuring label-sums in $m$ subsets $a_1, \ldots, a_m$ of $\{x_1, \ldots, x_n\}$. We call these subsets pools and assume that the vector $\sigma := (\sigma_1, \ldots, \sigma_n) \in \{0, 1, \ldots, d\}^n$ is chosen uniformly at random from all vectors containing exactly $k_i$ entries of value $i$ for each $i \in \{1, \ldots, d\}$. For each $i = 1, \ldots, d$, $k_i = \varepsilon_i n^\theta$ for $\varepsilon_i = \Theta(1)$ and $\theta \in (0, 1)$, and we abbreviate $k = \sum_{i=1}^d k_i = n^\theta$. Finally, $W = \sum_{i=1}^d i k_i$ denotes the total weight of $\sigma$.

In the following, we will represent pooling schemes $\{a_1, \ldots, a_m\}$ as bipartite multi-graphs. In this representation, the pools and the items yield the two vertex classes of the bipartite graph (see Figure 1). An edge in this graph is present whenever the incident item is an element of the incident pool. This prescription yields a multi-graph since in our scheme, items are allowed to appear multiple times in a given pool. We denote the neighbourhood of item $x_i$ in this graph by $\partial x_i$ and the neighbourhood of pool $a_j$ by $\partial a_j$. These are understood to be multi-sets of pools and items, respectively. If we work with sets rather than multi-sets, we use the notation $\partial^\star x_i, \partial^\star a_j$ for the sets obtained from $\partial x_i$ and $\partial a_j$. Finally, as in the introduction, we denote the label-sum of pool $a_i$ by $\hat{\sigma}_i$ such that $\hat{\sigma}_j = \sum_{i \in \partial a_j} \sigma_i$. Generally, if $z \in \mathbb{R}^p$, we also use the notation $z(t)$ the refer to the $t$-th component of $z$, where $t = 1, \ldots, p$.

## 2.2 The pooling scheme

The polynomial-time pooling scheme that constitutes the basis for our algorithm is based on the introduction of a "spatial" order to both items and pools. To this end, we first partition $V := \{x_1, \ldots, x_n\}$ into $\ell$ *compartments* $V[s], \ldots, V[\ell+s-1] \subset V$ of (almost) equal sizes $|V[i]| \in \{\lfloor n/\ell \rfloor, \lceil n/\ell \rceil\}$, where $s = \lceil \ell^{1/2-\varepsilon/2} \rceil$ and $\ell = \lceil k^{1/2-\varepsilon} \rceil$ for some fixed $\varepsilon > 0$. This partition will allow us to successively infer the labels of each compartment, where we proceed from 1 to $\ell$ and use the information of previous compartments along the way. However, to get this idea started properly, the first few compartments need some extra attention.

We facilitate the initial steps of the algorithm by the introduction of $s - 1$ artificial compartments $V[1], \ldots, V[s-1]$ (this also explains why the labelling in the previous paragraph starts at $s$). These contain $n' = (s-1)\lceil n/\ell \rceil$ *auxiliary items* that are distributed equally among the compartments. To equip these auxiliary items with labels that behave as the original variable labels, we then sample an assignment $\tau \in \{0, 1, \ldots, d\}^{n'}$ uniformly at random from all vectors with exactly $k_i' = \lceil (s-1)\varepsilon_i k \ell^{-1} \rceil$ items of weight $i$, $i = 1, \ldots, d$. This only takes polynomial time and in particular, $\tau$ is known. In the remainder of this article, we call the $s - 1$ compartments $V_{\text{seed}} = V[1], \ldots, V[s-1]$ the *seed*, while the remaining compartments constitute the *bulk* $V_{\text{bulk}}$.

Analogously to the partition of the bulk, for an integer $m$ divisible by $\ell + s - 1$, we divide the $m$ pools into $\ell + s - 1$ many compartments $F[1], \ldots, F[\ell + s - 1]$ such that each of the compartments contains exactly $m/(\ell + s - 1) \sim m/\ell$ pools.

After this partitioning, items for the pools are chosen as follows: First, let

$$\Gamma := \frac{ns}{\sqrt{m}(\ell + s - 1)} + O(s)$$

be an integer divisible by $s$, which will be the number of items in each pool. Let $j \in \{1, \ldots, \ell + s - 1\}$ be the index of one of the pool compartments. Then each pool $a \in F[j]$ chooses $\Gamma$ items exclusively from the $s$ "previous" compartments $V[j - (s-1)], \ldots, V[j]$, where it selects exactly $\Gamma/s$ items from each of these compartments uniformly at random with replacement. Here and in the following, for $r = 1, \ldots, s - 1$, we identify $V[-r]$ with $V[\ell + s - 1 - (r-1)]$ as well as $F[\ell + s - 1 + r]$ with $F[r]$, which equips the random bipartite graph with a ring structure. In particular, the number $s$ is called *sliding window*. The terminology and construction are illustrated by Figure 2.

As described above, this pooling scheme gives rise to a bipartite multi-graph. We denote the (random) degrees of the items $x_1, \ldots, x_n$ in this graph by $\Delta_{x_1}, \ldots, \Delta_{x_n}$. The number of neighbours of item $x_i$ among the elements of $F[i + j]$ is denoted by $\Delta_{x_i}[j]$, where $j = 0, \ldots, s - 1$. Furthermore, as the pools sample their items with replacement, we introduce $\Delta_{x_i}^{\star}, \Delta_{x_i}^{\star}[j]$ as the corresponding numbers of *distinct* pools that item $x_i$ participates in. As we will see later, the effect of multi-edges is almost negligible, as $\Delta_{x_i}^{\star} = (1 - o(1))\Delta_{x_i}$ w.h.p., but their presence facilitates the analysis. Finally, set $\Delta := \mathbb{E}[\Delta_{x_i}]$ and $\Delta^{\star} := \mathbb{E}\left[\Delta_{x_i}^{\star}\right]$.

## 3 ALGORITHM OUTLINE

*Motivation.* As a starting point for the algorithm, we consider the influence of an arbitrary variable $x \in V$ on the pools that it participates in.

While a change in the label of $x$ typically only marginally affects the outcome of the query at one of its neighbouring pools, the situation is different if we consider all $\Delta_x$ neighbours of $x$ jointly. This number is binomially distributed and thus concentrated around its mean $\Delta = \sqrt{m}s/\ell$. More precisely, the Chernoff bound guarantees that, w.h.p.,

$$\Delta_x = \Delta \pm \ln n \sqrt{\Delta} = (1 + o(1))\Delta.$$

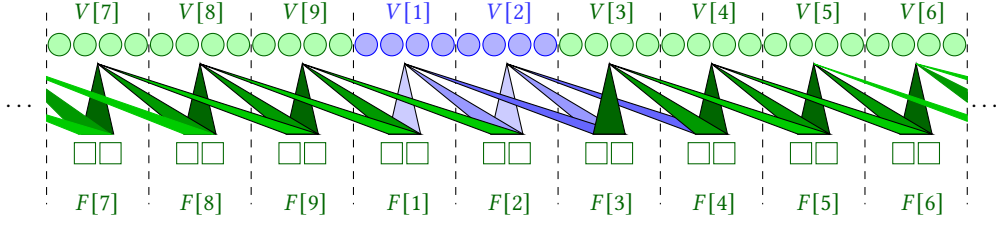

Fig. 2. Schematic representation of the pooling scheme with $n = 28$ items, $\ell = 7$ compartments, 18 measurements and a sliding window of size $s = 3$. The number of (blue) auxiliary items, whose weight is known a priory, is 8 in this example. The figure is an adaptation of Figure 2 in [5].

By definition of $s, \ell$ and for $m = \Omega(k)$ (in agreement with the information-theoretic lower bound), we have $\Delta = \Omega(k^{1/4})$. Therefore, if we define the *neighbourhood sum* of $x$ as the sum of all the outcomes of pools in which $x$ occurs, and alter the label of $x$, the change in the neighbourhood sum is of order $\Omega(\Delta)$. As a consequence, the neighbourhood sum of $x$ is highly dependent on $\sigma_x$.

In light of this observation, it is natural to try to discern the labels by thresholding neighborhood sums, provided that these are concentrated well enough for items of different types and that sufficiently many measurements are conducted. This idea, which does *not* make use of our sophisticated pooling scheme, has previously been applied to the quantitative group testing problem [12], where the authors show that it leads to successful recovery of $\sigma$ using $\Theta(k \ln n)$ measurements w.h.p.

We aim to improve this result through the spatially coupled design, which provides additional information on the labels at each inference step. For simplicity, we assume from now on that $x \in V[s]$. Then the pooling scheme in combination with the auxiliary compartments ensures that:

- Item $x$ is *only* contained in pools from the $s$ compartments $F[s], \ldots, F[2s-1]$.
- For each $j = 0, \ldots, s-1$, any pool from compartment $F[s+j]$ contains a proportion of $1 - (j+1)/s$ of already correctly inferred labels w.h.p. (namely, those from compartments $V[1], \ldots, V[s-1]$).

Therefore, rather than to simply sum up the labels of items that share a pool with $x$, one should construct the neighbourhood sum and then subtract all contributions from previously inferred labels in order to lay bare the influence of the unknown labels. We call this sum the *unexplained neighbourhood sum* of $x$.

Considering unexplained neighbourhood sums of items already yields some improvement over naive neighbourhood sums. However, the second observation above also indicates that pools from compartments close to $F[s]$ actually reveal *more* information on the items in $V[s]$ than pools from far apart compartments, since they contain a larger portion of known labels. For example, the pools in $F[s]$ have unexplained neighbourhood sums that exclusively involve labels from $V[s]$. One idea to incorporate this imbalance between the information coming from the different compartments and to further improve the concept of an unexplained neighbourhood sum is to introduce weights $\omega_1, \ldots, \omega_s \in (0, 1]$ to scale the contributions accordingly. We call such a compartment-wise linear combination of unexplained neighbourhood sums *weighted unexplained neighbourhood sum*[2].

Alas, it turns out that again, this process does not yield the desired improvement. This is due to the fact that while the information from close compartments is indeed much more valuable, on the other hand the expected size of the unexplained neighbourhood sum is by a factor of $s = n^{\Omega(1)}$ larger in the farthest away compartment than in the closest one. Thus, despite the contrary effort, the influence of the first compartment almost vanishes in the weighted unexplained neighbourhood sum. To compensate for this effect, we normalise the unexplained neighbourhood sum in each compartment and sum up those normalised quantities in the described weighted fashion with the *weighted normalised unexplained neighbourhood sum* $\mathcal{N}_x$, which is the core quantity of our algorithm.

---

[2]A weighted unexplained neighbourhood sum is the key quantity in the analysis of optimal binary group testing [5].

*Key quantities.* In this subsection, we formalise the notions of the last subsection. Let $\tilde{\sigma} \in \{0, \ldots, d\}^n$ be the current estimate of $\sigma$ during any step of the algorithm. Then for each item $x \in V[i]$, $i = s, \ldots, \ell + s - 1$, and $j = 0, \ldots, s - 1$ we first define the *unexplained neighbourhood sum* of $x$ into compartment $F[i+j]$ with respect to the estimate $\tilde{\sigma}$ as

$$\mathcal{U}_x^j := \mathcal{U}_x^j(\hat{\sigma}, \tilde{\sigma}) = \sum_{a \in \partial^\star x \cap F[i+j]} \left( \hat{\sigma}_a - \sum_{y \in \partial a} \tilde{\sigma}_y \right). \tag{3.1}$$

We illustrate this random variable for the special case $d = 1$. In this case, given the pool design,

$$\mathcal{U}_x^j(\hat{\sigma}, \sigma) \approx \text{Bin}\left( \frac{j+1}{s^2} \Delta_x^\star \Gamma, \frac{k}{n} \right) + \frac{\Delta_x}{s} \sigma_x.$$

Indeed, $x$ has roughly $\Delta_x^\star/s$ neighbours in compartment $F[i+j]$, of which each features roughly $(j+1)\Gamma/s$ so far unexplained items. Of course, this is only a heuristic approximation of the unexplained neighbourhood sum, but still instructive. The precise distribution of $U_x^j$ is derived in Lemma 3.2.

In a next step, we define the *normalised unexplained neighbourhood sum* $\mathcal{N}_x^j$ of $x$ into compartment $F[i+j]$. As explained in the previous paragraph, this quantity takes (an approximation of) the expectation and the variance of $\mathcal{U}_x^j$ into account. For this, we define a conditional estimate of the expectation of $\mathcal{U}_x^j$ by setting

$$M_x^j := \sum_{w=1}^d w \frac{k_w \ell}{n} \left( (j+1) \Delta_x^\star[j] \frac{\Gamma}{s} - \Delta_x[j] \right).$$

The variance is, with high probability, $(1 + o(1))(j+1)k^{2\varepsilon}$ by the choice of $\Gamma$ and the concentration properties of $\Delta_x^\star$. The normalised unexplained neighbourhood sum $\mathcal{N}_x^j$ of $x$ into compartment $F[i+j]$ is then defined as

$$\mathcal{N}_x^j := \frac{\mathcal{U}_x^j - M_x^j}{\sqrt{(j+1)k^{2\varepsilon}}}.$$

As $\mathcal{U}_x^j$, the quantity $\mathcal{N}_x^j$ depends on $\sigma_x$. Indeed, it is approximately of order

$$\mathcal{N}_x^j(\hat{\sigma}, \sigma) \approx \sigma_x \Delta s^{-1} \sqrt{(j+1)k^{2\varepsilon}}^{-1} = C \sigma_x \sqrt{j+1}^{-1}$$

for some constant $C > 0$. But, in contrast to $\mathcal{U}_x^j$, it is comparable between items in *close* compartments are *far apart* compartments.

Finally, the *weighted normalised unexplained neighbourhood sum* of item $x$ with a specified choice of weights is defined as

$$\mathcal{N}_x := \sum_{j=0}^{s-1} (j+1)^{-0.5} \mathcal{N}_x^j. \tag{3.2}$$

*Finding thresholds.* The idea of the inference algorithm is quite simple: We define thresholds $T^{0,1}, T^{1,2}, \ldots, T^{d-1,d}$ such that item $x$ is classified as having weight $i \in \{1, \ldots, d-1\}$ if $T^{i-1,i} < \mathcal{N}_x \leq T^{i,i+1}$, as having label 0 if $\mathcal{N}_x \leq T^{0,1}$ and as having label $d$ otherwise. With respect to the information-theoretic lower bound $m_{\text{QGT}}$ in the quantitative group testing problem, assume that the spatially coupled pooling scheme involves

$$m = 2c\frac{1-\theta}{\theta}k$$

pools for some constant $c = c_{d,\theta} \geq 1$ that may depend on $d$ and $\theta$. We then define the thresholds $T^{i,i+1}$ for $i \in \{1,\ldots,d-1\}$ and additionally $T^{0,1}$ as

$$T^{i,i+1} := \left(i + \frac{1}{2}\right)\sqrt{\frac{2c(1-\theta)}{\theta}}\ln s \quad \text{and} \quad T^{0,1} := \frac{1}{1+\sqrt{\theta}}\sqrt{\frac{2c(1-\theta)}{\theta}}\ln s.$$

The threshold $T^{0,1}$ takes a slightly form, as it tells apart the much more numerous items of weight 0 from the items of non-zero weight. The choice of $T^{0,1},\ldots,T^{d-1,d}$ is explained in the appendix.

### 3.1 Algorithm

Using the notation from the previous sections, our algorithm now does the following.

Set $\tilde{\sigma} = 0 \in \{0,1,\ldots,d\}^n$;
**for** $i = s,\ldots,\ell + s - 1$ **do**
  For any individual $x \in V[i]$ calculate $\mathcal{N}_x$;
  Set $\tilde{\sigma}_x = \begin{cases} 0, & \text{if } \mathcal{N}_x < T^{0,1} \\ d, & \text{if } \mathcal{N}_x \geq T^{d-1,d} \\ b, & \text{if } 1 \leq b \leq d-1 \text{ and } \mathcal{N}_x \in \left(T^{b-1,b}, T^{b,b+1}\right]. \end{cases}$
**end**

**Algorithm 1:** Algorithm for the pooled data problem with thresholds $T^{0,1},\ldots,T^{d-1,d}$.

It turns out that, for a decent constant $c$, we find the following performance guarantee for Algorithm 1.

PROPOSITION 3.1. *If* $c \geq \frac{(4+\delta)(1+\sqrt{\theta})}{1-\sqrt{\theta}}\sum_{w=1}^{d} w^2 \varepsilon_w$ *for some* $\delta > 0$, *the output* $\tilde{\sigma}$ *of Algorithm 1 coincides with* $\sigma$ *w.h.p.*

*Outline of the proof of Proposition 3.1.* We prove Proposition 3.1 in three steps. To begin with, we study the conditional distribution of the unexplained neighbourhood sums $\mathcal{N}_x$, given the following idealised information: the weight of $x$, the weightings in which $x$ participates as well as the correct numbers of items of weight $w$ in compartment $j$ for $w = 0,\ldots,d$ and $j \in [\ell]$. Secondly, we use the properties of this conditional distribution to bound the probabilities that thresholding $\mathcal{N}_x$ (had we access to it) would lead to a wrong classification of $x$. We then prove Proposition 3.1 by an inductive argument that ensures that with sufficiently high probability, $\mathcal{N}_x$ is not too far from $\mathbf{N}_x$ for all $x$.

The proof starts from the idea to find a reasonably well-behaved substitute of $\mathcal{N}_x$. This idea can be implemented by the observation that the quantity $\mathcal{U}_x^j(\hat{\sigma}, \tilde{\sigma})$ has a particularly accessible form if $\tilde{\sigma}$ agrees with the ground truth on the previously inferred compartments, and we introduce specific notation for the unexplained neighbourhood sum of $x$ into compartment $F[i+j]$ with respect to the correct labels $\sigma_1,\ldots,\sigma_{n(i-1)/\ell}$. To this end, denote by $\sigma^i$ the $n$-dimensional vector which agrees with $\sigma$ up to coordinate $n(i-1)/\ell$ and with $\sigma_b^i = 0$ for $b > n(i-1)/\ell$. We then define

$$U_x^j := \mathcal{U}_x^j(\hat{\sigma}, \sigma^i).$$

Thus, $U_x^j$ is the (random) unexplained neighbourhood sum of item $x$ into compartment $F[i+j]$ with respect to the true $\sigma$. Moreover, denote by $k_i^{(j)}$ the random number of items of weight $i$ in compartment $j$ in the spatial coupling set-up

and abbreviate

$$\underline{k}^{(j)} := \left(k_i^{(j)}\right)_{i=1\ldots d} \qquad \text{and} \qquad \underline{k} := \left(\underline{k}^{(j)}\right)_{j \in [\ell]}.$$

Finally, let the $\sigma$-algebra $\mathcal{E}_x$ be

$$\mathcal{E}_x := \sigma\left(\partial x, \sigma_x, \underline{k}\right).$$

With this, we can also define the associated quantities

$$N_x^j := \frac{U_x^j - \mathbb{E}\left[U_j^x \big| \mathcal{E}_x\right] + \Delta_x[j]\sigma_x}{\sqrt{(j+1)k^{2\varepsilon}}} \qquad \text{as well as} \qquad N_x := \sum_{j=0}^{s-1}(j+1)^{-0.5}N_x^j, \tag{3.3}$$

Of course, if the estimate $\tilde{\sigma}$ has inferred every label correctly so far, then $U_x^j$ agrees with $\mathcal{U}_x^j$. Unfortunately, the actual values of $\sigma$ and $U_x^j$ are unknown at any specific stage of the algorithm, and thus, it can neither compute $U_x^j$ nor $N_x$ exactly. However, the main strategy of the proof is to analyse $N_x$ and then to show that w.h.p., the guess $\mathcal{N}_x$ is sufficiently close to $N_x$ for the concentration properties of $N_x$ to be transferred.

More precisely, the proof of Proposition 3.1 proceeds in three steps. First, in Lemma 3.2, we analyse the distribution of each $U_x^j$ given $\mathcal{E}_x$, which is basically a weighted sum over the coordinates of independent multinomially distributed random vectors. Second, we use this distributional insight to also calculate the conditional means and variances of $U_x^j$ given $\mathcal{E}_x$ and conclude that conditionally on $\sigma_x = w$ for any $w = 0, \ldots, d$, the idealised weighted normalised unexplained neighbourhood sums $N_x$ are tightly concentrated around their means. Third, Proposition 3.1 will follow from an inductive argument based on the spatial coupling of the pools.

We start with the following distributional insight which formalises the intuition that was given in the last section for the case $d = 1$. A detailed proof can be found in the appendix.

LEMMA 3.2. *Let $x \in V[i]$ and $j \in \{0, \ldots, s-1\}$. Let*

$$X_x^{(i:j)} := \text{Mult}\left(\frac{\Delta_x^\star[j]\Gamma}{s} - \Delta_x[j], \frac{k_1^{(j)} - \mathbf{1}\{\sigma_x = 1\}}{n/\ell - 1}, \ldots, \ell\frac{k_d^{(j)} - \mathbf{1}\{\sigma_x = d\}}{n/\ell - 1}\right)$$

*and*

$$X_x^{(r:j)} \sim \text{Mult}\left(\frac{\Delta_x^\star[r]\Gamma}{s}, \frac{\ell k_1^{(r)}}{n}, \ldots, \frac{\ell k_d^{(r)}}{n}\right)$$

*for $r = i+1, \ldots, i+j$ be independent multinomial random variables given $\mathcal{E}_x$. Then*

$$U_x^j \stackrel{d}{=} \Delta_x[j]\sigma_x + \sum_{r=i}^{i+j}\sum_{w=1}^{d} w X_x^{(r:j)}(w) \qquad \text{given } \mathcal{E}_x. \tag{3.4}$$

As it turns out that $N_x$ given $\mathcal{E}_x$ can be written as a weighted sum over negatively associated Bernoulli random variables. Therefore, by a straightforward application of Bernstein's inequality, we find the following concentration property.

LEMMA 3.3. *For any $\alpha \in (0,1)$, set $T_\alpha^{0,1} := (1-\alpha)\sqrt{\frac{2c(1-\theta)}{\theta}\ln s}$. Then, there exists a constant $D > 0$ that depends on $d$ and $\theta$ and a sequence $\varepsilon_n = o\left(n^{-2}\right)$ such that*

$$\mathbb{P}\left(N_x > T_\alpha^{0,1}\big|\sigma_x = 0\right) \leq Ds^{-\frac{(1-\alpha)^2 c(1-\theta)}{\theta \sum_{w=1}^{d} w^2 \varepsilon_w}} + \varepsilon_n, \mathbb{P}\left(N_x \leq T_\alpha^{0,1}\big|\sigma_x = 1\right) \leq Ds^{-\frac{\alpha^2 c(1-\theta)}{\theta \sum_{w=1}^{d} w^2 \varepsilon_w}} + \varepsilon_n$$

as well as

$$\mathbb{P}\left(\mathcal{N}_x > T^{i,i+1} \middle| \sigma_x = i\right) \leq Ds^{-\frac{c(1-\theta)}{4\theta \sum_{w=1}^d w^2 \varepsilon_w}} + \varepsilon_n \quad \text{and}$$

$$\mathbb{P}\left(\mathcal{N}_x \leq T^{i,i+1} \middle| \sigma_x = i+1\right) \leq Ds^{-\frac{c(1-\theta)}{4\theta \sum_{w=1}^d w^2 \varepsilon_w}} + \varepsilon_n$$

for $i = 1, \ldots, d-1$.

A detailed verification of this claim is the content of Appendix B.3.

Finally, Proposition 3.1 follows from an inductive argument based on the spatial coupling of the pools. More precisely, for any given compartment $V[i]$, we condition on the event that the algorithm's estimate $\tilde{\sigma}$ agrees with $\sigma$ on all previous compartments $V[1], \ldots, V[i-1]$. In this case, for $x \in V[i]$, $\mathcal{N}_x^j$ and $\mathbf{N}_x^j$ differ only in their centering, and we argue that $\mathcal{N}_x^j$ is close to $\mathbf{N}_x^j$, so that w.h.p., also the quantities $\mathcal{N}_x$ are concentrated well enough in between the thresholds $T^{0,1}, \ldots, T^{d-1,d}$ to correctly infer all items from compartment $V[i]$. From a more quantitative point of view, a short but rather technical calculation shows that, with high probability, $\mathcal{N}_x = \mathbf{N}_x + \tilde{O}\left(\sqrt{\frac{s^2}{\ell}}\right)$, while $\mathbf{N}_x = \Theta(\ln s)$. By our choice of $\ell$ and $s$, the approximation error is small enough to be negligible. The proof of this assertion is carried out in detail in Appendix B.4.

### 3.2 Proof of Theorem 1.1

Theorem 1.1 now is an immediate consequence of Proposition 3.1, if we show that Algorithm 1 computes the estimate $\tilde{\sigma}$ on the bulk items in polynomial time. To this end, it is necessary to calculate both the quantities $\mathcal{N}_x$ as well as to threshold them. However, for each item $x \in V_{\text{bulk}}$, this can be done with $n^2$ elementary operations. Therefore, the running time is polynomial in the number of items.

∎

## 4 CONCLUSION

*On the spatially coupled pooling design.* For simplicity, we discuss our results in the quantitative group testing setting. Similar conclusions hold for more general $d = O(1)$. In the case $d = 1$, Corollary 1.2 states that our algorithm succeeds with high probability while using a number of queries that is within a constant factor of the number of queries required by *any* algorithm, even with unlimited computational power. This (small) gap can, in principle, have three reasons. Firstly, our pooling scheme might be sub-optimal. Secondly, the inference algorithm might not be optimal given the pooling scheme. Thirdly, there might be a gap between efficient and non-efficient algorithms.

To discuss these options in more detail, we briefly sketch that, at least for small $\theta$, we cannot hope to find a better inference algorithm on the chosen design. The argument resembles the information-theoretic counting bound from the introduction, and a similar line of thought was presented by Feige and Lellouche [10]. It makes use of the fact that the number of items with label 1 inside each pool is, approximately, $\text{Bin}(\Gamma, k/n)$ distributed. Therefore, with high probability, all results stem from an interval of length $\ln(n)k^{1/8}$ by Chernoff's bound. To distinguish the $\binom{n}{k}$ possible values of $\sigma$,

$$\left(\ln(n)k^{1/8}\right)^m \geq \binom{n}{k} \iff m \geq (8 + o(1))\frac{1-\theta}{\theta}k$$

measurements are required asymptotically. For small $\theta$, Algorithm 1 matches this bound. While this simple counting bound does not directly apply to the studied design due to the presence of multi-edges, the argument stays valid as, with high probability, every item is part of almost only different measurements. For larger values of $\theta$, an improvement might be achievable through a tolerance for few mistakes in the algorithm's classification and a *back-propagation* approach to repair them locally. This would require less strict concentration of the number of items with specific weight in the second neighbourhood of a specific given item.

As the algorithm is optimal for small $\theta$, the pooling design itself has to be sub-optimal. This evidence is further supported by the fact that the design would be information-theoretically optimal, if the underlying graph was much denser. More precisely, it is known that $\Gamma \sim n/2$ and $\Delta \sim m/2$ allows reconstruction of $\sigma$ w.h.p. by exhaustive search at $m_{\text{non-ada}}^{\text{QGT}}$ [12, 22]. Unfortunately, we cannot make the pooling design denser due to our inference algorithm. As the number of items with weight 1 in each compartment has to be estimated by $k\ell^{-1}$, those estimates yield approximation errors with respect to the correct underlying distribution which become more severe as the graph becomes denser. The pooling design was chosen as dense as possible such that those approximation errors are, with high probability, negligible in comparison to the difference of items with label zero and label one in the neighborhood sum. A detailed verification can be found in Appendix C.

*Future directions.* We present an efficient algorithm that closes the previous multiplicative $\ln n$ gap between the simple information-theoretic lower bound and all previously known efficient algorithms for the pooled data problem and its special cases. A natural question with respect to the quantitative group testing problem is whether there is a pooling design coming with an efficient inference algorithm that matches the known exponential time constructions up to the correct constant. However, it seems unlikely that our ideas can be stretched much further onto this point. Rather, we think that one needs to come up with new ideas for a pooling scheme or a modification of the inference algorithm in order to achieve this goal.

## ACKNOWLEDGMENTS

Max Hahn-Klimroth is supported by DFG CO 646/5. Noela Müller is supported by ERC-Grant 772606-PTRCSP. We thank Dominik Kaaser and Philipp Loick for fruitful discussions on the quantitative group testing problem. We furthermore thank an anonymous referee for bringing up a simplification of the spatially coupled pooling design.

## A  PRELIMINARIES

In Section A.1, we fix a parametrisation for the hypergeometric distribution. For the reader's convenience, we also list some well-known concentration inequalities as references. Section A.2 collects some properties of the typical degrees in our spatial coupling scheme which follow immediately from the statements in Section A.1.

### A.1  General

We denote by $\mathrm{Hyp}(N, M, K)$ the *hypergeometric distribution* with parameters $N \in \mathbb{N}_0, M, K \in \{0, \ldots, N\}$, which is defined by its point masses

$$\mathrm{Hyp}(N, M, K)(\{j\}) = \frac{\binom{M}{j}\binom{N-M}{K-j}}{\binom{N}{K}} \qquad \text{for} \quad j \in \{\max\{0, K+M-N\}\}, \ldots, \min\{K, M\}.$$

LEMMA A.1 (CHERNOFF BOUND FOR THE BINOMIAL DISTRIBUTION). *Let $X \sim \mathrm{Bin}(n, p)$. Then for any $\varepsilon > 0$,*

$$\mathbb{P}(X \geq (1+\varepsilon)\mathbb{E}[X]) \leq \exp\left(-\frac{\varepsilon^2}{2+\varepsilon}\mathbb{E}[X]\right) \qquad \text{and}$$

$$\mathbb{P}(X \leq (1-\varepsilon)\mathbb{E}[X]) \leq \exp\left(-\frac{\varepsilon^2}{2}\mathbb{E}[X]\right).$$

LEMMA A.2 (CHERNOFF BOUND FOR THE HYPERGEOMETRIC DISTRIBUTION [14]). *Let $X$ be a $Hyp(N, M, K)$-distributed random variable and let $t \geq 0$. Then*

$$\mathbb{P}(X - \mathbb{E}[X] \geq t) \leq \exp\left(-\frac{t^2}{2(KM/N + t/3)}\right) \quad \text{and}$$

$$\mathbb{P}(X - \mathbb{E}[X] \geq -t) \leq \exp\left(-\frac{t^2}{2KM/N}\right).$$

PROPOSITION A.3 (BERNSTEIN INEQUALITY). *Let $X_1, \ldots, X_n$ be independent random variables such that $\mathbb{E}[X_i] = 0$ and $|X_i| \leq z$ almost surely for all $i \in [n]$ and a constant $z > 0$. Moreover, let $\sigma^2 := \frac{1}{n}\sum_{i=1}^n \text{Var}(X_i)$. Then for all $\varepsilon > 0$,*

$$\mathbb{P}\left[\sum_{i=1}^n X_i \geq \varepsilon n\right] \leq \exp\left(-\frac{n\varepsilon^2}{2\sigma^2 + 2z\varepsilon/3}\right).$$

*Definition A.4 (Negative association). Let $X = (X_1, \ldots, X_d)$ be a random vector. Then the family of random variables $X_1, \ldots, X_d$ is said to be negatively associated if for every two disjoint index sets $I, J \subseteq [d]$ we have*

$$\mathbb{E}\left[f(X_i : i \in I)g(X_j : j \in J)\right] \leq \mathbb{E}\left[f(X_i : i \in I)\right]\mathbb{E}\left[g(X_j : j \in J)\right]$$

*for all functions $f : \mathbb{R}^{|I|} \to \mathbb{R}$ and $g : \mathbb{R}^{|J|} \to \mathbb{R}$ that are either both non-decreasing or both non-increasing.*

COROLLARY A.5 (COROLLARY EC.4 OF CHEN ET AL. [4]). *Bernstein's inequality (Proposition A.3) holds for negatively associated random variables.*

## A.2 Properties of the pooling scheme

This section provides bounds on the degrees of the items in the spatially coupled pooling scheme as well as on the numbers of items of a specific weight per compartment. We start with the degrees of the items. For this, let $i \in \{1, \ldots, \ell + s - 1\}$ and $x \in V[i]$, $j = 0, \ldots, s - 1$. Recall from section 2.2 that $\Delta_x[j]$ denotes the number of pools $a \in F[i + j]$ such that $x \in \partial a$, where multiple occurrences are counted multiple times. On the other hand, $\Delta_x^\star[j]$ is the number of *distinct* pools from compartment $F[i + j]$ in which $x$ participates. Furthermore, we defined

$$\Delta_x = \sum_{j=0}^{s-1} \Delta_x[j], \qquad \Delta = \mathbb{E}[\Delta_x] \quad \text{as well as} \quad \Delta_x^\star = \sum_{j=0}^{s-1} \Delta_x^\star[j], \qquad \Delta^\star = \mathbb{E}[\Delta_x^\star].$$

COROLLARY A.6. *Suppose that $m = \Theta(n^\theta)$. With probability $1 - o(n^{-2})$, for all $i \in \{1, \ldots, \ell\}$ and all $j \in \{0, \ldots, s-1\}$,*

$$\frac{\Delta}{s} - \ln(n)\sqrt{\frac{\Delta}{s}} \leq \min_{x \in V[i]} \Delta_x[j] \leq \max_{x \in V[i]} \Delta_x[j] \leq \frac{\Delta}{s} + \ln(n)\sqrt{\frac{\Delta}{s}}$$

*as well as*

$$\frac{\Delta^\star}{s} - \ln n\sqrt{\frac{\Delta^\star}{s}} \leq \min_{x \in V[i]} \Delta_x^\star[j] \leq \max_{x \in V[i]} \Delta_x^\star[j] \leq \frac{\Delta^\star}{s} + \ln n\sqrt{\frac{\Delta^\star}{s}}.$$

PROOF. Each pool $a \in F[i + j]$ chooses $\Gamma/s = n/(\sqrt{m}\ell) + O(1)$ items in compartment $V[i]$ uniformly at random, and thus $\Delta_x[j] \sim \text{Bin}\left(n\sqrt{m}/\ell^2 + O(m/\ell), \ell/n + O(\ell^2/n^2)\right)$. Thus, $\mathbb{E}[\Delta_x[j]] = \sqrt{m}/\ell + O(m/n) = \Delta/s + O(n^{\theta-1})$ and by Lemma A.1,

$$\mathbb{P}\left(\Delta_x[j] - \frac{\Delta}{s} \geq \sqrt{\frac{\Delta}{s}}\ln n\right) \leq \exp\left(-\Theta\left(\ln^2 n\right)\right)$$

as well as

$$\mathbb{P}\left(\Delta_x[j] - \frac{\Delta}{s} \leq -\sqrt{\frac{\Delta}{s}} \ln n\right) \leq \exp\left(-\Theta\left(\ln^2 n\right)\right).$$

The corollary now follows from a union bound over all items $x$ and neighbouring compartments $j$. □

COROLLARY A.7. *Suppose that $m = \Theta(n^\theta)$. We have that $\Delta = \Delta^\star(1 + o(1))$.*

PROOF. For all $i \in [n]$, $\Delta_i \sim \text{Bin}\left(m\Gamma/\ell, \ell/n + O(\ell^2/n^2)\right)$, so $\Delta = m\Gamma/n + O(\Gamma m\ell/n^2) = \sqrt{m}s/\ell + O(sm/n)$. On the other hand, for all $i \in [n]$,

$$\Delta_i^\star \sim \text{Bin}\left(ms/\ell, 1 - \left(1 - \ell/n + O(\ell^2/n^2)\right)^{\Gamma/s}\right).$$

Using $\Gamma/s = n/(\sqrt{m}\ell) + O(1)$, Bernoulli's inequality (first $\leq$) and $1 + x \leq e^x$ in combination with $e^x \leq 1 + x + x^2/2$ for $x \leq 0$ (second $\leq$), we arrive at

$$1 - \frac{1}{\sqrt{m}} - O\left(\frac{\ell}{n}\right) \leq \left(1 - \frac{\ell}{n} + O\left(\frac{\ell^2}{n^2}\right)\right)^{\Gamma/s} \leq 1 - \frac{1}{\sqrt{m}} + O\left(\frac{1}{m}\right) + O\left(\frac{\ell}{n}\right).$$

This yields $\Delta^\star = \frac{\sqrt{m}s}{\ell}\left(1 + O\left(\frac{1}{\sqrt{m}}\right) + O\left(\frac{\ell\sqrt{m}}{n}\right)\right)$ and thus the corollary. □

The preceding corollaries show that in fact, the number of pools in which each item participates, counted with or without multiplicity, is concentrated around $\Delta/s$. We finally study concentration properties of the numbers of items of a specific weight per compartment. For this, recall the definition of $k_j^{(i)}$ as the number of items of weight $j = 0, \ldots, d$ in compartment $V[i]$ from Section 3.1.

COROLLARY A.8. *Suppose that $m = \Theta(n^\theta)$. With probability $1 - o(n^{-2})$ we find for $j = 1, \ldots, d$ that*

$$\frac{k_j}{\ell} - \ln n \sqrt{\frac{k_j}{\ell}} \leq \min_{i \in [\ell]} k_j^{(i)} \leq \max_{i \in [\ell]} k_j^{(i)} \leq \frac{k_j}{\ell} + \ln n \sqrt{\frac{k_j}{\ell}}.$$

PROOF. Since the ground truth $\sigma$ is uniformly distributed among all $\sigma \in \{0, 1, \ldots, d\}^n$ with exactly $k_j$ entries of value $j$, for each $j = 0, \ldots, d$, $k_j^{(i)} \sim \text{Hyp}(n, k_j, |V[i]|)$. The assertion again follows from a combination of the Chernoff (Lemma A.2) and the union bound.

□

# B PROOF OF PROPOSITION 3.1

Recall that $k_0, \ldots, k_d$ denote the numbers of items of weight $0, \ldots, d$, respectively, and that $k_i = \varepsilon_i k$ for $i \in [d]$, where $\varepsilon_i = \Theta(1)$. Moreover, denote by $k_i^{(j)}$ the random number of items of weight $i$ in compartment $j$ in the spatial coupling set-up and abbreviate

$$\underline{k}^{(j)} := \left(k_i^{(j)}\right)_{i=1\ldots d} \quad \text{and} \quad \underline{k} := \left(\underline{k}^{(j)}\right)_{j \in [\ell]}.$$

Let $x$ be any item of the bulk. We first describe the conditional distribution of the unexplained neighbourhood sum with respect to $\sigma$. To this end, let

$$\mathcal{E}_x := \sigma\left(\partial x, \sigma_x, \underline{k}\right).$$

Lemma 3.2 asserts that Given $\mathcal{E}_x$, the conditional distribution of $U_x^j$ takes the simple form as a weighted sum over the components of independent multinomially distributed vectors.

## B.1 Proof of Lemma 3.2

Recall that $U_x^j$ is the unexplained neighbourhood sum of individual $x$ in compartment $i + j$ under perfect knowledge of the weights in the first $i - 1$ compartments. More precisely, there are $\Delta_x^\star[j]$ different weightings in the $(i + j)$th compartment that contribute to $U_x^j$, each of which itself comprises contributions of items from compartments $V[r]$, $r = i + j - s + 1, \ldots, i + j$. However, in $U_x^j$, only the contributions by items from compartments $V[r], r = i, \ldots, i + j$ are counted. Each of these $i + j + 1$ compartments contributes $\Gamma/s$ (non-necessarily distinct) items to each of the $\Delta_x^\star[j]$ tests. We may thus write

$$U_x^j = \sum_{r=i}^{i+j} \sum_{w=1}^{d} w \sum_{a \in \partial^\star x \cap F[i+j]} \sum_{y=1}^{\Gamma/s} \mathbf{1}\,\{\text{Neighbour number } y \text{ of } a \text{ in } V[r] \text{ has weight } w\}. \tag{B.1}$$

Given $\mathcal{E}_x$, the precise numbers $k_0^{(r)}, \ldots, k_d^{(r)}$ of coins of each weight in compartment $r \in [\ell]$ are known. Moreover, each weighting chooses its coins independently with replacement from compartment $r$. The vector of the numbers of coins of weights $0, \ldots, d$ chosen from compartment $r$ by a neighbour $a$ of $x$ in $F[i + j]$ therefore follows a multinomial distribution. The parameters of this distribution slightly depend on whether $r = i$ or $r > i$, as compartment $i$ contains $x$. For $r = i$, given $\mathcal{E}_x$, there is a deterministic contribution of $\Delta_x[j]\sigma_x$ to $U_x^j$ that is solely due to $x$. The remaining contribution of compartment $i$ is by items with weight distribution

$$\text{Mult}(\Delta_x^\star[j]\Gamma/s - \Delta_x[j]; k_0^{(i)} - \mathbf{1}\,\{\sigma_x = w\}, \ldots, k_d^{(i)} - \mathbf{1}\,\{\sigma_x = w\}),$$

as each weighting chooses its participants independently and uniformly, given that it does not choose $x$. For $r > i$, the total contribution to (B.1) from $V[r]$ is by items with weights distribution $\text{Mult}(\Delta_x^\star[j]\Gamma/s; k_0^{(r)}, \ldots, k_d^{(r)})$. Moreover, given $\mathcal{E}_x$, all these multinomial random variables are independent, and the lemma follows.

## B.2 The conditional unexplained neighbourhood sum

From Lemma 3.2, the conditional expectations and variances of $U_x^j$ and $N_x^j$ are immediate. For brevity, let

$$n_x^{(r:j)} := \Delta_x^\star[j]\frac{\Gamma}{s} - \mathbf{1}\,\{r = i\}\,\Delta_x[j] \quad \text{and} \quad p_x^{(r:j)}(w) := \frac{k_w^{(j)} - \mathbf{1}\,\{r = i\}\,\sigma_x}{\frac{n}{\ell} - \mathbf{1}\,\{r = i\}} \tag{B.2}$$

for $r = i, \ldots, i + j$ and $w = 0, \ldots, d$ denote the parameters of the multinomial distributions in Lemma 3.2.

COROLLARY B.1. *Let $x \in V[i]$ and $j = 0, \ldots, s - 1$. Then*

$$\mathbb{E}\left[U_x^j \middle| \mathcal{E}_x\right] = \Delta_x[j]\sigma_x + \sum_{r=i}^{i+j} \sum_{w=1}^{d} w n_x^{(r:j)} p_x^{(r:j)}(w) \quad \text{and} \tag{B.3}$$

$$\text{Var}\left(U_x^j \middle| \mathcal{E}_x\right) = \sum_{r=i}^{i+j} n_x^{(r:j)} \left(\sum_{w=1}^{d} w^2 p_x^{(r:j)}(w) - \left(\sum_{w=1}^{d} w p_x^{(r:j)}(w)\right)^2\right).$$

*The normalised random variables $N_x^j$ correspondingly satisfy*

$$\mathbb{E}\left[N_x^j \middle| \mathcal{E}_x\right] = \frac{\Delta_x[j]\sigma_x}{\sqrt{(j+1)k^{2\varepsilon}}} \quad \text{and} \quad \text{Var}\left(N_x^j \middle| \mathcal{E}_x\right) = \frac{1}{(j+1)k^{2\varepsilon}}\text{Var}\left(U_x^j \middle| \mathcal{E}_x\right).$$

Corollary B.1 provides the basis for the definition of the thresholds $T^{0,1}, T^{1,2}, \ldots, T^{d-1,d}$ by stating the explicit dependence of $\mathbb{E}[N_x^j|\mathcal{E}_x]$ on $\sigma_x$. To obtain a first quantitative understanding of the typical sizes of $N_x^j$ and $N_x$ given

$\{\sigma_x = w\}$, $w = 0, \ldots, d$, recall the quantities

$$\Gamma = \frac{ns}{\sqrt{m\ell}}, \qquad \Delta = \frac{\sqrt{m}s}{\ell}, \qquad \ell = k^{1/2-\varepsilon} \quad \text{and} \quad s = k^{1/4-\varepsilon}.$$

The number of conducted measurements $m$ will be parametrised as

$$m = 2c_d \frac{1-\theta}{\theta} k, \tag{B.4}$$

for constants $c_d > 0$ such that the constant $c_d$ is the ratio of $m$ to the information-theoretic lower bound in the case $d = 1$. Corollary B.1 and $\mathbb{E}[\Delta_x[j]] = \Delta/s$ then imply that

$$\mathbb{E}\left[\mathbb{E}\left[N_x^j \big| \mathcal{E}_x\right] \big| \sigma_x = w\right] = w\sqrt{\frac{2c_d(1-\theta)}{(j+1)\theta}} \quad \text{and}$$

$$\mathbb{E}\left[\mathbb{E}\left[N_x \mid \mathcal{E}_x\right] \big| \sigma_x = w\right] = w\sqrt{\frac{2c_d(1-\theta)}{\theta}} \sum_{j=1}^{s} \frac{1}{j} = w\sqrt{\frac{2c_d(1-\theta)}{\theta}} \ln s + O(1)$$

for $w = 0, \ldots, d$. In other words, the difference in the conditional expectations of $\mathbb{E}[N_x^j|\mathcal{E}_x]$ given $\sigma_x = w$ and given $\sigma_x = w'$ is of order $\ln s$. Our strategy to define thresholds $T^{i,i+1}$ for $i = 0, \ldots, d-1$ such that we classify $x$ as having weight 0 if $\mathcal{N}_x \leq T^{0,1}$, as having weight $i \in [d-1]$ if $T^{i-1,i} < \mathcal{N}_x \leq T^{i,i+1}$, and as having weight $d$ otherwise is based on the idea that these means are sufficiently far apart and large deviations of the random variables from these means are sufficiently unlikely to correctly infer the weights. We thus define the thresholds $T^{i,i+1}$ for $i \in [d-1]$ as the average of the corresponding two consecutive expectations,

$$T^{i,i+1} := \left(i + \frac{1}{2}\right)\sqrt{\frac{2c_d(1-\theta)}{\theta}} \ln s.$$

However, the case $i = 0$ needs to be treated differently, as it tells apart the much more numerous coins of weight 0 from the coins of non-zero weight. This will become clearer in the next sections. Let $\alpha \in (0, 1)$. We then set

$$T_\alpha^{0,1} := (1-\alpha)\sqrt{\frac{2c_d(1-\theta)}{\theta}} \ln s.$$

This ansatz will be further specified by an explicit choice of $\alpha$ in the following two subsections, where we prove Proposition 3.1. In Section B.3 we derive the necessary concentration results for the random variables $N_x$ conditionally on $\{\sigma_x = w\}$. Using an inductive argument implemented by the spatial coupling, we then show in Section B.4 that even though we have no access to $N_x$ from a practical point of view, our statistics $\mathcal{N}_x$ is sufficiently close to it to infer the correct weights w.h.p.

### B.3 Concentration

PROOF OF LEMMA 3.3. It is an immediate consequence of Lemma 3.2 that conditionally on $\mathcal{E}_x$, we have the distributional identity

$$N_x \stackrel{d}{=} \sum_{j=0}^{s-1} \left( \frac{\Delta_x[j]\sigma_x}{(j+1)k^\varepsilon} + \sum_{r=i}^{i+j} \sum_{w=1}^{d} w \frac{\text{Mult}\left(n_x^{(r:j)}; p_x^{(r:j)}\right)(w) - n_x^{(r:j)} p_x^{(r:j)}(w)}{(j+1)k^\varepsilon} \right),$$

as a weighted sum over centered multinomial coordinates. Now fix a compartment $r = 0, \ldots, s-1$ and consider the corresponding multinomial distribution $\text{Mult}(\boldsymbol{n}_x^{(r:j)}; \boldsymbol{p}_x^{(r:j)})$. Given $\mathcal{E}_x$, let

$$B_{x,t,w}^{(r:j)} \stackrel{d}{=} \text{Be}\left(\boldsymbol{p}_x^{(r:j)}(w)\right), \qquad w = 0, \ldots, d, \ t \in [\boldsymbol{n}_x^{(r:j)}].$$

Then $(B_{x,t,w}^{(r:j)})_{w \in \{0,\ldots,d\}, t \in [\boldsymbol{n}_x^{(r:j)}]}$ is a negatively associated family of random variables with

$$\text{Mult}\left(\boldsymbol{n}_x^{(r:j)}; \boldsymbol{p}_x^{(r:j)}\right)(w) \stackrel{d}{=} \sum_{t=1}^{\boldsymbol{n}_x^{(r:j)}} B_{x,t,w}^{(r:j)}$$

for all $w = 0, \ldots, d$. Moreover, the families

$$(B_{x,t,w}^{(0:j)})_{w \in \{0,\ldots,d\}, t \in [\boldsymbol{n}_x^{(r:j)}]}, \ldots, (B_{x,t,w}^{(s-1:j)})_{w \in \{0,\ldots,d\}, t \in [\boldsymbol{n}_x^{(r:j)}]}$$

are independent and thus conditionally on $\mathcal{E}_x$ and on a more granular level,

$$\boldsymbol{N}_x \stackrel{d}{=} \sum_{j=0}^{s-1} \left( \frac{\Delta_x[j]\sigma_x}{(j+1)k^\varepsilon} + \sum_{r=i}^{i+j} \sum_{w=1}^{d} \sum_{t=1}^{\boldsymbol{n}_x^{(r:j)}} w \frac{B_{x,t,w}^{(r:j)} - \boldsymbol{p}_x^{(r:j)}(w)}{(j+1)k^\varepsilon} \right).$$

This decomposition shows that $\boldsymbol{N}_x | \mathcal{E}_x$ is distributed as a constant term plus a sum of $d \sum_{j=0}^{s-1}((j+1)\Delta_x^\star[j]\Gamma/s - \Delta_x[j])$ independent, centered random variables, each of which is almost surely bounded by $dk^{-\varepsilon}$ thanks to our scaling. Thus, by Bernstein's inequality for negatively associated random variables (Proposition A.5),

$$\mathbb{P}\left(\boldsymbol{N}_x - \sum_{j=0}^{s-1} \frac{\Delta_x[j]\sigma_x}{(j+1)k^\varepsilon} \geq T \Big| \mathcal{E}_x\right) \leq \exp\left(-\frac{T^2}{2\text{Var}(\boldsymbol{N}_x | \mathcal{E}_x) + 2Tdk^{-\varepsilon}/3}\right) \quad \text{a.s.} \tag{B.5}$$

for $T > 0$. This conditional estimate allows us to upper bound the probability that the unexplained neighbourhood sum exceeds a given threshold $T$, given that item $x$ has weight $i = 0, \ldots, d$, as

$$\mathbb{P}(\boldsymbol{N}_x \geq T | \sigma_x = i) = \mathbb{E}\left[\mathbb{P}(\boldsymbol{N}_x \geq T | \mathcal{E}_x) | \sigma_x = i\right]$$

$$\leq \mathbb{E}\left[\exp\left(-\frac{\left(T - \sum_{j=0}^{s-1} \frac{\Delta_x[j]i}{(j+1)k^\varepsilon}\right)^2}{2\text{Var}(\boldsymbol{N}_x | \mathcal{E}_x) + 2Tk^{-\varepsilon}/3}\right) \Big| \sigma_x = i\right].$$

The random variables $\sum_{j=0}^{s-1} \frac{\Delta_x[j]i}{(j+1)k^\varepsilon}$ and $\text{Var}(\boldsymbol{N}_x | \mathcal{E}_x)$ can be bounded in the following way. Define the concentration events

$$\mathcal{K}_1 := \bigcap_{i=1}^{\ell} \bigcap_{j=1}^{s} \bigcap_{x \in V[i]} \left\{ \left|\Delta_x[j] - \frac{\Delta}{s}\right| \leq 2\ln n \sqrt{\frac{\Delta}{s}} \right\},$$

$$\mathcal{K}_2 := \bigcap_{i=1}^{\ell} \bigcap_{j=1}^{s} \bigcap_{x \in V[i]} \left\{ \left|\Delta_x^\star[j] - \frac{\Delta}{s}\right| \leq 2\ln n \sqrt{\frac{\Delta}{s}} \right\} \quad \text{and}$$

$$\mathcal{K}_3 := \bigcap_{i=1}^{\ell} \bigcap_{w=0}^{d} \left\{ \left|k_w^{(i)} - \frac{k_w}{\ell}\right| \leq 2\ln n \sqrt{\frac{k_w}{\ell}} \right\}.$$

With this, we set

$$\mathcal{K} := \mathcal{K}_1 \cap \mathcal{K}_2 \cap \mathcal{K}_3.$$

Due to Corollaries A.6 and A.8, we know that $\mathcal{K}$ holds with probability $1 - o(n^{-2})$. On $\mathcal{K}$, there exists a constant $A_\theta > 0$ such that for $i \in [d-1]$

$$\left(T^{i,i+1} - \sum_{j=0}^{s-1} \frac{\Delta_x[j]i}{(j+1)k^\varepsilon}\right)^2 \geq \frac{1}{2}\sqrt{\frac{2c_d(1-\theta)}{\theta}} \ln s - A_\theta \quad \text{as well as}$$

$$\left(T_\alpha^{0,1} - \sum_{j=0}^{s-1} \frac{\Delta_x[j]i}{(j+1)k^\varepsilon}\right)^2 \geq (1-\alpha)\sqrt{\frac{2c_d(1-\theta)}{\theta}} \ln s.$$

Similarly, on $\mathcal{K}$, for $n$ large enough,

$$\text{Var}\left(\mathbf{N}_x | \mathcal{E}_x\right) \leq \left(\sum_{w=1}^{d} w^2 \varepsilon_w\right) \ln s - A_\theta.$$

For $n$ large enough, with probability $1 - o(n^{-2})$, $\text{Var}\left(\mathbf{N}_x | \mathcal{E}_x\right) \leq \ln s + 1$. Due to Corollaries A.6 and A.8, we thus have

$$\mathbb{P}\left(\mathbf{N}_x \geq T^{i,i+1} | \boldsymbol{\sigma}_x = i\right) \leq \exp\left(-\frac{c_d(1-\theta)}{4\theta \sum_{w=1}^{d} w^2 \varepsilon_w} \ln s + B_{\theta,d}\right) + o\left(n^{-2}\right)$$

$$\leq C_\theta s^{-\frac{c_d(1-\theta)}{4\theta \sum_{w=1}^{d} w^2 \varepsilon_w}} + o\left(n^{-2}\right)$$

and

$$\mathbb{P}\left(\mathbf{N}_x \geq T_\alpha^{0,1} | \boldsymbol{\sigma}_x = 0\right) \leq \exp\left(-\frac{(1-\alpha)^2 c_d(1-\theta)}{\theta \sum_{w=1}^{d} w^2 \varepsilon_w} \ln s + B_{\theta,d}\right) + o\left(n^{-2}\right)$$

$$\leq C_\theta s^{-\frac{(1-\alpha)^2 c_d(1-\theta)}{\theta \sum_{w=1}^{d} w^2 \varepsilon_w}} + o\left(n^{-2}\right).$$

Similarly,

$$\mathbb{P}\left(\mathbf{N}_x \leq T^{i,i+1} | \boldsymbol{\sigma}_x = i+1\right) \leq \exp\left(-\frac{c_d(1-\theta)}{4\theta \sum_{w=1}^{d} w^2 \varepsilon_w} \ln s + B_{\theta,d}\right) + o\left(n^{-2}\right)$$

$$\leq C_\theta s^{-\frac{c_d(1-\theta)}{4\theta \sum_{w=1}^{d} w^2 \varepsilon_w}} + o\left(n^{-2}\right)$$

and

$$\mathbb{P}\left(\mathbf{N}_x \leq T_\alpha^{0,1} | \boldsymbol{\sigma}_x = 1\right) \leq \exp\left(-\frac{\alpha^2 c_d(1-\theta)}{\theta \sum_{w=1}^{d} w^2 \varepsilon_w} \ln s + B_{\theta,d}\right) + o\left(n^{-2}\right)$$

$$\leq C_\theta s^{-\frac{\alpha^2)^2 c_d(1-\theta)}{\theta \sum_{w=1}^{d} w^2 \varepsilon_w}} + o\left(n^{-2}\right).$$

□

## B.4 Proof of Proposition 3.1

PROOF OF PROPOSITION 3.1. Throughout this proof, we fix the choice $\alpha = \sqrt{\theta}/(1 + \sqrt{\theta})$ as in the statement of Proposition 3.1. Moreover, as $m > m_{\text{PD}}$, we also have

$$f := \frac{c_d \cdot (1/4 - \varepsilon) \cdot (1 - \sqrt{\theta})}{(1 + \sqrt{\theta}) \cdot (\sum_{w=1}^{d} w^2 \varepsilon_w)} > 1. \tag{B.6}$$

Set $\mathcal{V}_i := V[1] \cup \ldots \cup V[i]$ for $i \in [\ell]$. We prove by induction on $i \in \{s, \ldots \ell - 1\}$ that there exists a constant $E > 0$ such that

$$\left| \mathbb{P}\left[ \tilde{\sigma}|_{\mathcal{V}_i} = \sigma|_{\mathcal{V}_i} \right] - \mathbb{P}\left[ \tilde{\sigma}|_{\mathcal{V}_{i+1}} = \sigma|_{\mathcal{V}_{i+1}} \right] \right| \le E \left( \frac{n^{1-f}}{\ell} + \frac{k^{1-f}}{\ell} \right).$$

We discuss the base case $i = s$ in detail and then state the necessary modifications for the inductive step that follows along the same lines. As all seed item's labels are known, we have $\mathbb{P}\left[ \tilde{\sigma}|_{\mathcal{V}_s} = \sigma|_{\mathcal{V}_s} \right] = 1$. For $n$ large enough, we may therefore condition on the event $\{\tilde{\sigma}|_{\mathcal{V}_s} = \sigma|_{\mathcal{V}_s}\}$ which implies that

$$\mathbb{P}\left[ \tilde{\sigma}|_{\mathcal{V}_{s+1}} = \sigma|_{\mathcal{V}_{s+1}} \right] = \mathbb{P}\left[ \text{for all } x \in V[s+1] : \tilde{\sigma}_x = \sigma_x \middle| \tilde{\sigma}|_{\mathcal{V}_s} = \sigma|_{\mathcal{V}_s} \right] \mathbb{P}\left[ \tilde{\sigma}|_{\mathcal{V}_s} = \sigma|_{\mathcal{V}_s} \right]$$

$$\ge \left( 1 - \sum_{x \in V[s+1]} \mathbb{P}\left[ \tilde{\sigma}_x \ne \sigma_x \middle| \tilde{\sigma}|_{\mathcal{V}_s} = \sigma|_{\mathcal{V}_s} \right] \right) \mathbb{P}\left[ \tilde{\sigma}|_{\mathcal{V}_s} = \sigma|_{\mathcal{V}_s} \right]. \tag{B.7}$$

Let now $x \in V[s+1]$. Distinguishing by the true weight of $x$, we have

$$\mathbb{P}\left[ \tilde{\sigma}_x \ne \sigma_x \middle| \tilde{\sigma}|_{\mathcal{V}_s} = \sigma|_{\mathcal{V}_s} \right] = \mathbb{P}\left( \mathcal{N}_x \ge T_\alpha^{0,1}, \sigma_x = 0 \middle| \tilde{\sigma}|_{\mathcal{V}_s} = \sigma|_{\mathcal{V}_s} \right)$$

$$+ \sum_{w=1}^{d-1} \mathbb{P}\left( \mathcal{N}_x \ge T^{w,w+1}, \sigma_x = w \middle| \tilde{\sigma}|_{\mathcal{V}_s} = \sigma|_{\mathcal{V}_s} \right)$$

$$+ \mathbb{P}\left( \mathcal{N}_x \le T_\alpha^{0,1}, \sigma_x = 1 \middle| \tilde{\sigma}|_{\mathcal{V}_s} = \sigma|_{\mathcal{V}_s} \right)$$

$$+ \sum_{w=2}^{d} \mathbb{P}\left( \mathcal{N}_x \le T^{w-1,w}, \sigma_x = w \middle| \tilde{\sigma}|_{\mathcal{V}_s} = \sigma|_{\mathcal{V}_s} \right).$$

On $\tilde{\sigma}|_{\mathcal{V}_s} = \sigma|_{\mathcal{V}_s}$, we have $\tilde{\mathcal{U}}_x^j = U_x^j$ for all $j = 0, \ldots, s-1$, and thus

$$\mathcal{N}_x = N_x + \sum_{j=0}^{s-1} \frac{1}{(j+1)k^{\varepsilon}} \left( \mathbb{E}\left[ U_x^j | \mathcal{E}_x \right] - \Delta_x[j]\sigma_x - M_x^j \right) =: N_x + R_x.$$

Thus

$$\mathbb{P}\left[\tilde{\sigma}_x \neq \sigma_x, \tilde{\sigma}|_{\mathcal{V}_s} = \sigma|_{\mathcal{V}_s}\right] = \frac{n-k}{n}\mathbb{P}\left(N_x + R_x \geq T_\alpha^{0,1}|\sigma_x = 0\right) \tag{B.8}$$

$$+ \sum_{w=1}^{d-1} \frac{k_w}{n}\mathbb{P}\left(N_x + R_x \geq T^{w,w+1}|\sigma_x = w\right)$$

$$\tag{B.9}$$

$$+ \frac{k_1}{n}\mathbb{P}\left(N_x + R_x \leq T_\alpha^{0,1}|\sigma_x = 1\right) \tag{B.10}$$

$$+ \sum_{w=2}^{d} \frac{k_w}{n}\mathbb{P}\left(N_x + R_x \leq T^{w-1,w}|\sigma_x = w\right) \tag{B.11}$$

and therefore, since the above probabilities are the same for each $x$, plugging (B.8) into (B.7) yields

$$0 \leq \mathbb{P}\left[\tilde{\sigma}|_{\mathcal{V}_s} = \sigma|_{\mathcal{V}_s}\right] - \mathbb{P}\left[\tilde{\sigma}|_{\mathcal{V}_{s+1}} = \sigma|_{\mathcal{V}_{s+1}}\right]$$

$$\leq \frac{n-k}{\ell}\mathbb{P}\left(N_x + R_x \geq T_\alpha^{0,1}|\sigma_x = 0\right) + \sum_{w=1}^{d-1} \frac{k_w}{\ell}\mathbb{P}\left(N_x + R_x \geq T^{w,w+1}|\sigma_x = w\right)$$

$$+ \frac{k_1}{\ell}\mathbb{P}\left(N_x + R_x \leq T_\alpha^{0,1}|\sigma_x = 1\right) + \sum_{w=2}^{d} \frac{k_w}{\ell}\mathbb{P}\left(N_x + R_x \leq T^{w-1,w}|\sigma_x = w\right).$$

Now, the point is that $R_x$ is small. Let the event $\mathcal{K}$ be defined as in the proof of Lemma 3.3. Then on $\mathcal{K}$, for $n$ large enough

$$|R_x| \leq 20 \sum_{w=1}^{d} w\sqrt{\varepsilon_w} k^{-\varepsilon/2} \ln n.$$

Therefore by Lemma 3.3, there exists a constant $D$ that only depends on $d$ and $\theta$ such that

$$\mathbb{P}\left[\tilde{\sigma}|_{\mathcal{V}_i} = \sigma|_{\mathcal{V}_i}\right] - \mathbb{P}\left[\tilde{\sigma}|_{\mathcal{V}_{i+1}} = \sigma|_{\mathcal{V}_{i+1}}\right]$$

$$\leq D\frac{n-k}{\ell}s^{-\frac{c_d(1-\theta)(1-\alpha)^2}{\theta \sum_{w=1}^{d} w^2 \varepsilon_w}} + 2D\left(\sum_{w=1}^{d} w\varepsilon_w\right)\frac{k}{\ell}s^{-\frac{c_d(1-\theta)}{4\theta \sum_{w=1}^{d} w^2 \varepsilon_w}}$$

$$+ \quad D\frac{k_1}{\ell}s^{-\frac{c_d(1-\theta)\alpha^2}{\theta \sum_{w=1}^{d} w^2 \varepsilon_w}} + o\left(\frac{1}{n\ell}\right).$$

This looks rather complicated, but using the definition of $s = k^{1/4-\varepsilon}$ and (B.6) of $f$, we arrive at

$$\mathbb{P}\left[\tilde{\sigma}|_{\mathcal{V}_i} = \sigma|_{\mathcal{V}_i}\right] - \mathbb{P}\left[\tilde{\sigma}|_{\mathcal{V}_{i+1}} = \sigma|_{\mathcal{V}_{i+1}}\right]$$

$$\leq D\frac{n-k}{\ell}n^{-f} + 2D\left(\sum_{w=1}^{d} w\varepsilon_w\right)\frac{k}{\ell}k^{-f\frac{(1+\sqrt{\theta})^2}{4\theta}} + D\varepsilon_1\frac{k}{\ell}k^{-f} + o\left(\frac{1}{n\ell}\right).$$

In particular, as $\frac{(1+\sqrt{\theta})^2}{4\theta} > 1$ for all $\theta \in (0,1)$, there exists a constant $E > 0$ such that

$$\mathbb{P}\left[\tilde{\sigma}|_{\mathcal{V}_i} = \sigma|_{\mathcal{V}_i}\right] - \mathbb{P}\left[\tilde{\sigma}|_{\mathcal{V}_{i+1}} = \sigma|_{\mathcal{V}_{i+1}}\right] \leq E\left(\frac{n^{1-f}}{\ell} + \frac{k^{1-f}}{\ell}\right).$$

Finally, the inductive step $i \to i+1$ proceeds exactly along the same lines, where we use that due to the induction hypothesis,

$$\mathbb{P}\left[\tilde{\sigma}|_{\mathcal{V}_i} = \sigma|_{\mathcal{V}_i}\right] \geq \mathbb{P}\left(\tilde{\sigma}|_{\mathcal{V}_s} = \sigma|_{\mathcal{V}_s}\right) - \sum_{r=s}^{i-1} \mathbb{P}\left[\tilde{\sigma}|_{\mathcal{V}_r} = \sigma|_{\mathcal{V}_r}\right] - \mathbb{P}\left[\tilde{\sigma}|_{\mathcal{V}_{r+1}} = \sigma|_{\mathcal{V}_{r+1}}\right]$$

$$\geq \mathbb{P}\left(\tilde{\sigma}|_{\mathcal{V}_s} = \sigma|_{\mathcal{V}_s}\right) - E \sum_{r=s}^{i-1} \left(\frac{n^{1-f}}{\ell} + \frac{k^{1-f}}{\ell}\right) > 0$$

for $n$ large enough. Finally, Proposition 3.1 follows as above:

$$\mathbb{P}(\tilde{\sigma} = \sigma) = \mathbb{P}\left(\tilde{\sigma}|_{\mathcal{V}_s} = \sigma|_{\mathcal{V}_s}\right) - \sum_{i=s}^{\ell-1} \mathbb{P}\left[\tilde{\sigma}|_{\mathcal{V}_i} = \sigma|_{\mathcal{V}_i}\right] - \mathbb{P}\left[\tilde{\sigma}|_{\mathcal{V}_{i+1}} = \sigma|_{\mathcal{V}_{i+1}}\right]$$

$$\geq \mathbb{P}\left(\tilde{\sigma}|_{\mathcal{V}_s} = \sigma|_{\mathcal{V}_s}\right) - E\left(n^{1-f} + k^{1-f}\right) + o\left(\frac{1}{n}\right) = 1 - o(1).$$

□

REMARK B.2 (THE CHOICE OF $\alpha$ AND $c_d$). *In the proof of Proposition 3.1, we used $\alpha = \sqrt{\theta}/(1+\sqrt{\theta})$ in the first threshold. This choice can be justified by the following considerations: In the proof, we need to ensure that both*

$$n \cdot s^{-\frac{c_d(1-\theta)(1-\alpha)^2}{\theta \sum_{w=1}^{d} w^2 \varepsilon_w}} = o(1) \quad \text{as well as} \quad ks^{-\frac{c_d(1-\theta)\alpha^2}{\theta \sum_{w=1}^{d} w^2 \varepsilon_w}} = o(1) \tag{B.12}$$

*to guarantee the correctness of the algorithm. Simplifying both conditions yields*

$$\alpha < 1 - \sqrt{\frac{\sum_{w=1}^{d} w^2 \varepsilon_w}{(1/4-\varepsilon)c_d(1-\theta)}} \quad \text{as well as} \quad \alpha > \sqrt{\frac{\theta \sum_{w=1}^{d} w^2 \varepsilon_w}{(1/4-\varepsilon)c_d(1-\theta)}}.$$

*Now, first of all, in order for at least one $\alpha$ with these properties to exist we need*

$$1 - \sqrt{\frac{\sum_{w=1}^{d} w^2 \varepsilon_w}{(1/4-\varepsilon)c_d(1-\theta)}} - \sqrt{\frac{\theta \sum_{w=1}^{d} w^2 \varepsilon_w}{(1/4-\varepsilon)c_d(1-\theta)}} \geq 0,$$

*and simplifying yields the restriction*

$$c_d \geq \frac{1+\sqrt{\theta}}{1-\sqrt{\theta}} \cdot \frac{\sum_{w=1}^{d} w^2 \varepsilon_w}{1/4-\varepsilon}.$$

*Secondly, for the minimal choice of $c_d = \frac{1+\sqrt{\theta}}{1-\sqrt{\theta}} \cdot \frac{\sum_{w=1}^{d} w^2 \varepsilon_w}{1/4-\varepsilon}$, there is only the choice*

$$\alpha = \frac{1}{2}\left(1 - \frac{\sum_{w=1}^{d} w^2 \varepsilon_w}{(1/4-\varepsilon)c_d}\right) = \frac{\sqrt{\theta}}{1+\sqrt{\theta}}.$$

*Of course, if $c_d$ is larger than the lower bound, $\sqrt{\theta}/(1+\sqrt{\theta})$ is still a valid choice in the proof of Proposition 3.1.*

## C ON THE DENSITY OF THE POOLING DESIGN

This appendix sketches why the parameters of the spatially coupled design are chosen as they are. More precisely, we show that for denser pooling designs, the approximation errors within the inference algorithm become too large. For the sake of simplicity, we make the assumption $d = 1$.

We observe that Lemma 3.2 implies for a specific item $x \in V[i]$ that, given $\mathcal{E}_x$,

$$U_x^j \stackrel{d}{=} \Delta_x[j]\sigma_x + \mathrm{Bin}\left(\Delta_x^\star[j]\frac{\Gamma}{s} - \Delta_x[j], \frac{k^{(i)} - \sigma_x}{\frac{n}{\ell} - 1}\right) + \sum_{r=i+1}^{i+j} \mathrm{Bin}\left(\Delta_x^\star[j]\frac{\Gamma}{s}, \frac{k^{(r)}}{n/\ell}\right).$$

In the end, the algorithm would like to compute $\sum_{j=0}^{s-1} \frac{1}{\sqrt{j+1}}\left(U_x^j - \mathbb{E}\left[U_x^j \mid \mathcal{E}_x\right]\right)((j+1)k^{2\varepsilon})^{-\frac{1}{2}}$ but the best it can do is to estimate the parameter $\ell\frac{k^{(r)}}{n}$ by $\frac{k}{n}$ and estimate the expectation by $M_x^j$ based on this parameter. Those estimations are, due to the concentration properties of the number of items of weight one in each compartment, quite well, but nevertheless, they yield an approximation error of

$$\sum_{r=0}^{s-1}\frac{1}{\sqrt{r+1}}\frac{U_x^r - M_x^r}{\sqrt{(j+1)k^{2\varepsilon}}} = \sum_{r=0}^{s-1}\frac{1}{\sqrt{r+1}}\frac{U_x^r - \mathbb{E}\left[U_x^j \mid \mathcal{E}_x\right]}{\sqrt{(j+1)k^{2\varepsilon}}} + \sum_{r=0}^{s-1}\frac{1}{\sqrt{r+1}}\frac{\mathbb{E}\left[U_x^j \mid \mathcal{E}_x\right] - M_x^r}{\sqrt{(j+1)k^{2\varepsilon}}}.$$

As, ultimately, the algorithm needs to notice a deviation of $\Theta(\ln s)$, the errors due to the algorithm's approximation need to be of much lower order.

We find in first order that, with high probability, $\left|\mathbb{E}\left[U_x^r \mid \mathcal{E}_x\right] - M_x^r\right| = \tilde{\Theta}\left(j\frac{\Delta^\star \Gamma k}{s^2 n}\frac{k}{n}\frac{\ell}{k}\right)$ and therefore, with high probability,

$$\sum_{r=0}^{s-1}(r+1)^{-0.5}\frac{\left|\mathbb{E}\left[U_x^r \mid \mathcal{E}_x\right] - M_x^r\right|}{\sqrt{(j+1)k^{2\varepsilon}}} = \tilde{\Theta}\left(\frac{\sqrt{\Delta\Gamma\ell}}{\sqrt{n}}\right). \tag{C.1}$$

To take care of any poly-logarithmic factors, we therefore require $\Delta\Gamma\ell = n^{1-\Omega(1)}$ such that this approximation error vanishes. As $\Gamma = \Delta n/m$, this is equivalent to $\Delta^2 \ell m^{-1} = n^{-\Omega(1)}$. Recall that $m = \Theta(k)$ and that, by construction, we require $\Delta s^{-1} = n^{\Omega(1)}$ to ensure the concentration properties of the number of edges incident to an item in a specific compartment. Therefore, in total, setting $\Delta = \sqrt{m}\frac{s}{\ell}$, the conditions read

$$\Delta s^{-1} = \Theta\left(\sqrt{k}\ell^{-1}\right) = n^{\Omega(1)} \iff \ell = k^{1/2-\Omega(1)}$$

and

$$\Delta^2 \ell m^{-1} = s^2 \ell^{-1} = n^{-\Omega(1)} \iff s = \ell^{-1/2-\Omega(1)}.$$

Indeed, if we plug this into (C.1) we verify

$$\sum_{r=0}^{s-1}(r+1)^{-0.5}\frac{\left|\mathbb{E}\left[U_x^r \mid \mathcal{E}_x\right] - M_x^r\right|}{\sqrt{(j+1)k^{2\varepsilon}}} = \tilde{\Theta}\left(\sqrt{\Delta\Gamma\ell n^{-1}}\right) = \tilde{\Theta}\left(\sqrt{s^2\ell^{-1}}\right) = o(1)$$

by the choice of the parameters.

Observe that this approximation error carries over to the case $d = O(1)$ directly as can be seen as follows. As $k_w = \Theta(k)$, we find that the order of magnitude of each summand does not change. Furthermore, as there is only a constant factor more single summands, the asymptotically analysis stays valid.